\documentclass[conference]{IEEEtran}
\IEEEoverridecommandlockouts

\usepackage{cite}
\usepackage{amsmath,amssymb,amsfonts}
\usepackage{algorithmic}
\usepackage{multirow}
\usepackage{graphicx}
\usepackage{textcomp}
\usepackage{xcolor}
\usepackage{booktabs}
\usepackage{url}
\usepackage{pdfpages}
\def\BibTeX{{\rm B\kern-.05em{\sc i\kern-.025em b}\kern-.08em
    T\kern-.1667em\lower.7ex\hbox{E}\kern-.125emX}}

\makeatletter
\def\endthebibliography{%
  \def\@noitemerr{\@latex@warning{Empty `thebibliography' environment}}%
  \endlist
}
\makeatother

\newcommand{\todo}[1]{}
\renewcommand{\todo}[1]{{\color{red} TODO: {#1}}}
\begin{document}

\title{
Target Speaker ASR with Whisper\\
\thanks{\IEEEauthorrefmark{1}These two authors contributed equally.}
}

\author{\IEEEauthorblockN{
Alexander Polok\IEEEauthorrefmark{1}\IEEEauthorrefmark{2}, Dominik Klement\IEEEauthorrefmark{1}\IEEEauthorrefmark{2}\IEEEauthorrefmark{4}, Matthew Wiesner\IEEEauthorrefmark{4}, Sanjeev Khudanpur\IEEEauthorrefmark{4}, Jan Černocký\IEEEauthorrefmark{2}, Lukáš Burget\IEEEauthorrefmark{2}}
\IEEEauthorblockA{\IEEEauthorrefmark{2}Brno University of Technology, Brno, Czech Republic}
\IEEEauthorblockA{\IEEEauthorrefmark{4}Johns Hopkins University, Baltimore, United States of America}
}

\maketitle

\begin{abstract}
We propose a novel approach to enable the use of large, single-speaker ASR models, such as Whisper, for target speaker ASR. The key claim of this method is that it is much easier to model \emph{relative} differences among speakers by learning to condition on frame-level diarization outputs than to learn the space of all speaker embeddings. We find that adding even a single bias term per diarization output type before the first transformer block can transform single-speaker ASR models into target-speaker ASR models.
Our approach also supports speaker-attributed ASR by sequentially generating transcripts for each speaker in a diarization output. This simplified method outperforms baseline speech separation and diarization cascade by 12.9\,\% absolute ORC-WER on the NOTSOFAR-1 dataset.
\end{abstract}


\section{Introduction}
\label{sec:intro}

Self-supervised models~\cite{chen2022wavlm, hsu2021hubert, pratap2024scaling}, LLMs~\cite{achiam2023gpt, touvron2023llama}, and Whisper-style supervised models~\cite{radford2023robust, peng2023owsm} have demonstrated that scaling up models by using more parameters and extremely large amounts of data can enable the development of accurate automatic speech recognition (ASR) systems, even in relatively challenging environments.
However, these models have primarily been used in single-speaker, single-channel ASR systems, whereas most conversations involve multiple speakers and are often recorded by one or more microphones.

Approaches to handling this scenario generally integrate multiple blocks performing source separation, speaker segmentation, overlapped speech detection, post-hoc speaker clustering, and ASR in order to produce speaker-attributed conversation transcripts.
Alternatively, there are end-to-end systems transcribing multi-speaker speech directly using special tokens or multiple heads~\cite{raj23_surt2,li2023adaptingmultilingualasrmodels, Kanda2020SerializedOT,QIAN20181}.
Another approach is the use of a semi-end-to-end system, known as target-speaker ASR (TS-ASR)~\cite{Kanda2019_spkloss, zhang_23_conformer, ma2024extending, Zhang2023, Zili23_adapting, meng24c_interspeech} processing the original input mixture and transcribing each speaker individually.  

The conventional TS-ASR framework extracts speaker embeddings corresponding to the target speakers and uses them as an auxiliary input to the ASR system~\cite{karafiat2011ivector, Zili23_adapting}. While pre-trained speaker embedding extractors~\cite{dehak2010front,snyder2018x,wang2023wespeaker} can guide the ASR system by highlighting relevant information in the input and filtering out irrelevant content, they inherently require the system to learn how to map speaker embeddings to ASR speech embeddings.  
More recent methods include the use of adaptation layers and soft prompts to modify existing ASR models to incorporate speaker embeddings~\cite{ma2024extending} or speaker enrollment audios~\cite{meng24c_interspeech}. However, since these models are often trained on simulated datasets—due to the scarcity of multi-speaker ASR datasets and the need for a large number of speaker identities—they tend to experience significant performance degradation when applied to real-world multi-speaker scenarios~\cite{vinnikov24_interspeech, Mccowan2025_ami, Yu2021M2MetTI}.  

\begin{figure}[ht]
    \centering
    \includegraphics[width=0.9\columnwidth]{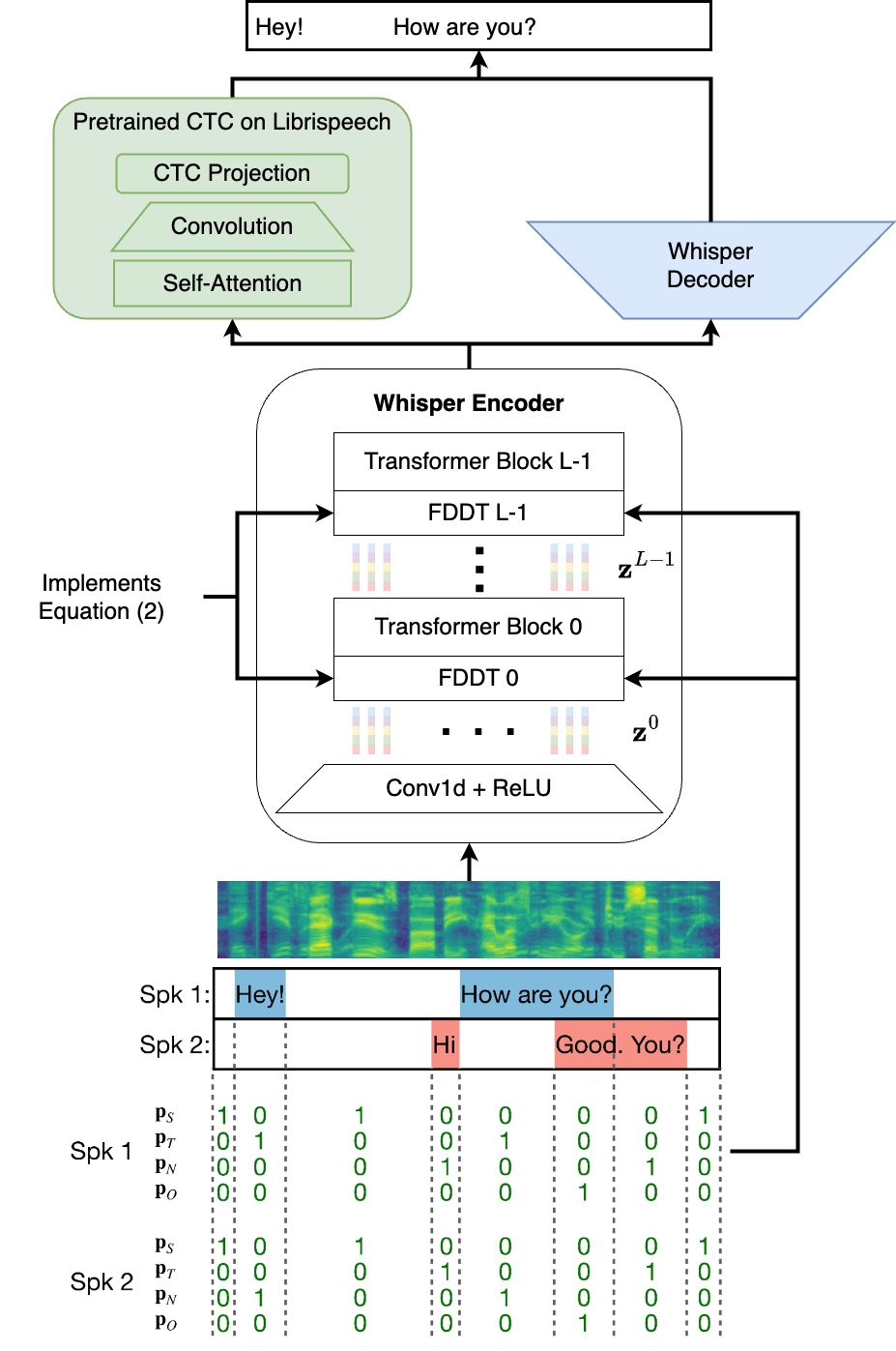}
    \caption{Proposed Diarization-Conditioned Whisper model: An input audio segment with multiple speakers is conditioned by frame-level diarization outputs $\begin{bmatrix} p_{\mathcal{S}}^t & p_{\mathcal{T}}^t & p_{\mathcal{N}}^t & p_{\mathcal{O}}^t \end{bmatrix}^T$ for each STNO class at every frame $t$. Affine transformations, applied to intermediate input representations $\mathbf{z}^l_{1:T}$, generate new embeddings, where $l$ is the layer index. The final frame-level embedding is a convex combination of these embeddings for each frame.}
\label{fig:ctc_whisper}
\end{figure}
In this paper, we propose a semi-end-to-end approach to TS-ASR that uses Whisper in a new way. Unlike previous TS-ASR methods, our system does not rely on speaker embeddings, but instead conditions directly on frame-level diarization outputs. We believe that, compared to the aforementioned embedding-based approaches, only "relative" differentiation between speakers is needed; the TS-ASR system does not need to adapt to an existing subspace of speaker embeddings. Training our model on labeled examples of both target and non-target speech may also improve speaker discrimination and improve robustness to diarization errors.

To validate our approach, we fine-tune Whisper models on the NOTSOFAR-1~\cite{vinnikov24_interspeech}, AMI~\cite{Mccowan2025_ami}, and Libri2Mix~\cite{cosentino2020librimixopensourcedatasetgeneralizable} datasets using ground truth speaker segmentation. 
Unless stated otherwise, ground truth segmentation is also used during inference. All experiments follow the NOTSOFAR-1 Challenge guidelines\footnote{\url{https://www.chimechallenge.org/challenges/chime8/task2/}}.

\section{Diarization-Conditioned Whisper} \label{sec:dcw}

This section presents the Diarization-Conditioned Whisper, a model built upon the Whisper architecture, designed to perform TS-ASR by conditioning on frame-level diarization outputs. 
An overview of the proposed model is shown in Fig.~\ref{fig:ctc_whisper}. We adapt Whisper for TS-ASR by adding Frame-Level Diarization Dependent Transformations (FDDT) modules, described in Section~\ref{sec:fddt}. 
These modules transform the model's internal representations in order to differentiate between the target- and non-target speakers in the audio.

\subsection{Silence, Target, Non-Target, and Overlap Masks}
\label{sec:stno_mask}

Let $\mathbf{D} \in [0,1]^{S \times T}$, where $S$ is the number of speakers in the recording, and $T$ is the number of frames,
represent the diarization output, with $d(s, t)$ denoting the probability that speaker $s$ is active at time frame $t$.

The dependency on the number of speakers in $\mathbf{D}$ can be a limiting factor for easily incorporating this mask into the model. 
To address this, let $s_k$ represent the target speaker. 
We define a distribution over the following mutually exclusive events for a frame at time $t$:

\begin{itemize}
    \item ${\mathcal{S}}$: Silence at time $t$.
    \item ${\mathcal{T}}$: Only the target speaker $s_k$ is active at $t$.
    \item ${\mathcal{N}}$: Non-target speaker(s) ($s \neq s_k$) active; target $s_k$ inactive at $t$.
    \item ${\mathcal{O}}$: Target $s_k$ and at least one non-target ($s \neq s_k$) active, causing overlap at $t$.
\end{itemize}

The probabilities of these events occurring at time frame $t$ can be calculated as:
\begin{align}
    p_{\mathcal{S}}^t  &= \prod_{s=1}^S (1 - d(s, t)), \quad 
    p_{\mathcal{T}}^t  = d(s_k, t)  \prod_{\substack{s=1 \\ s \neq s_k}}^S (1 - d(s, t)) \nonumber \\
    p_{\mathcal{N}}^t  &= \left(1 - p_{\mathcal{S}}^t\right) - d\left(s_k, t\right), \quad
    p_{\mathcal{O}}^t  = d(s_k, t) - p_{\mathcal{T}}^t
\end{align}

This definition allows us to use a fixed-sized target-speaker-dependent STNO (Silence, Target, Non-target, Overlap) mask $ \begin{bmatrix} p_{\mathcal{S}}^t & p_{\mathcal{T}}^t & p_{\mathcal{N}}^t & p_{\mathcal{O}}^t \end{bmatrix}^{\top}$.

\subsection{Input Masking}
\label{sec:mask}
Having the STNO mask, a straightforward way to perform target speaker ASR is to mask the signal by multiplying each frame by the probability that it is target speech or that it involves overlap with the target speaker.
Hence, if the target speaker is not active, the audio signal is set to 0 (i.e., silence).
We add $p_\mathcal{T}^t$ and $p_\mathcal{O}^t$ to ensure that both target speech and overlapping speech are preserved in the masked signal.
However, similar to source separation approaches, this method has limitations. It can introduce artifacts because we are creating a modified version of the input signal, and errors in diarization can propagate through the system, potentially affecting the model's performance.

\subsection{Frame-Level Diarization Dependent Transformations}
\label{sec:fddt}
To overcome issues of Input Masking, we designed a soft version called Frame-Level Diarization Dependent Transformations (FDDT). This approach modifies the frame-by-frame model inputs based on the diarization outputs.

Let $\mathbf{Z}^l \in \mathbb{R}^{d_{{m}} \times T}$ represent the frame-by-frame inputs to the $l$-th (Transformer) layer. We transform these hidden representations by applying four affine STNO layer- and class-specific transformations: $\mathbf{W}_{\mathcal{S}}^l, \mathbf{W}_{\mathcal{T}}^l, \mathbf{W}_{\mathcal{N}}^l, \mathbf{W}_{\mathcal{O}}^l \in \mathbb{R}^{d_{{m}} \times d_{{m}}}$ together with biases $\mathbf{b}_{\mathcal{S}}^l, \mathbf{b}_{\mathcal{T}}^l, \mathbf{b}_{\mathcal{N}}^l, \mathbf{b}_{\mathcal{O}}^l \in \mathbb{R}^{d_{m}}$ to obtain new speaker-specific hidden representations $\hat{\mathbf{Z}}^l = [\hat{\mathbf{z}}^l_1, \ldots, \hat{\mathbf{z}}^l_T]$ as
\begin{align}
\label{eq:FDDT}
\hat{\mathbf{z}}^l_t = &\left( \mathbf{W}_{\mathcal{S}}^l \mathbf{z}^l_t + \mathbf{b}_{\mathcal{S}}^l \right) p^t_{\mathcal{S}} + 
\left( \mathbf{W}_{\mathcal{T}}^l \mathbf{z}^l_t + \mathbf{b}_{\mathcal{T}}^l \right) p^t_{\mathcal{T}}  \nonumber \\
 &+ \left( \mathbf{W}_{\mathcal{N}}^l \mathbf{z}^l_t + \mathbf{b}_{\mathcal{N}}^l\right) p^t_{\mathcal{N}} + 
\left( \mathbf{W}_{\mathcal{O}}^l \mathbf{z}^l_t + \mathbf{b}_{\mathcal{O}}^l \right) p^t_{\mathcal{O}},
\end{align}

These transformations generate four distinct representations of the frame-by-frame inputs, each highlighting one of the STNO classes. A target-speaker-specific representation is formed by a convex combination of these terms, with weights derived from the STNO mask. The same transformation is applied to all frames with identical STNO masks.

Fine-tuning the model using randomly initialized FDDT matrices could easily disrupt the internal representations of the model.
Therefore, we propose initialization strategies to mitigate this risk: 
\begin{itemize}
    \item \textit{Identity Initialization} (Non-Disturbing Init): Here, biases are initialized with zero vectors, and weights are initialized as identity matrices. This method ensures that the model's internal representations are not altered.
    \item \textit{Suppressive Initialization}: To bias the model toward masking other speakers, we initialize the  $\mathbf{W}_{\mathcal{S}}^l, \mathbf{W}_{\mathcal{N}}^l$  weights as diagonal matrices with values close to zero, e.g., 0.1. This approach helps the model to distinguish between different types of speech, reinforcing the separation between the STNO classes. 
\end{itemize}

\section{Experiments}
\label{sec:experiments}
We primarily conducted our experiments on the new NOTSOFAR-1 dataset~\cite{vinnikov24_interspeech}, which includes 280 meetings, each averaging 6 minutes, capturing diverse real-world acoustic conditions and conversational dynamics.
To assess generalization and competitiveness, we also evaluated our best models on the synthetic Libri2Mix dataset~\cite{cosentino2020librimixopensourcedatasetgeneralizable} and on real-world meeting dataset AMI~\cite{Mccowan2025_ami}.
Experiments are divided into two parts: In Section~\ref{sec:ab_fddt}, we analyze the FDDT behavior under different weight structure constraints, initializations, numbers of additional parameters, and provided information. In Section~\ref{sec:ab_scaling}, we evaluate the framework’s performance as more parameters and data are added.

Source codes and recipes\footnote{\url{https://github.com/BUTSpeechFIT/TS-ASR-Whisper}} are built on top of the transformers library~\cite{wolf-etal-2020-transformers}. All models are evaluated with the Time-Constrained Optimal Reference Combination Word Error Rate (tcORC WER)~\cite{Neumann2023MeetEval}, referred to as WER throughout the text.

\begin{table}[t]
    \centering
    \caption{%
Comparison of the proposed system built with updated Whisper-large-v3-turbo~\cite{polok2024dicowdiarizationconditionedwhispertarget} alongside various multi-talker ASR systems. The top section includes systems where no additional ground truth information about speaker identity or segmentation is provided. The bottom section features models that directly or indirectly utilize ground truth segmentation information. Results marked with $\dagger$ are evaluated on utterance groups. Proposed ORC\,WER results marked with $\star$ were approximated by increasing the collar for tcORC\,WER.
    }\label{tab:SOTA_comp}
    \setlength{\tabcolsep}{3pt} 
    \small{\begin{tabular}{lccc}
        \toprule
         & \multicolumn{1}{p{1.5cm}}{\centering AMI-sdm \\test\\ORC\,WER}  & \multicolumn{1}{p{2cm}}{\centering NOTSOFAR-1 \\ eval-small \\ tcORC\,WER}   & \multicolumn{1}{p{1.5cm}}{\centering Libri2Mix \\ test-both\\ORC\,WER} \\
        \midrule
        Raj et al.~\cite{raj23_surt2} & $44.6^{\dagger}$ & $60.9^{\dagger}$ &  \\
        Vinnikov et al.~\cite{vinnikov24_interspeech} & & $35.5$ & \\
        Niu et al.~\cite{niu24_chime}  &  &17.7 & \\
        Proposed       & $18.0^\star$ & $22.6$ & 14.9 \\
        \midrule
        Input masking & 79.1 & 76.6 & 56.7 \\
        Ma et al.~\cite{ma2024extending} &  &  & 26.4  \\
        Zhang et al.~\cite{Zhang2023} &  &  & 23.5  \\
        Proposed & $16.5^\star$  & $19.1$  & 10.9 \\
        \bottomrule
    \end{tabular}}
\end{table}

\subsection{Training details}
\label{sec:training_details}
To enhance Whisper's performance, 
we incorporated an additional CTC (Connectionist Temporal Classification) head, following the hybrid CTC-attention-based training scheme proposed in~\cite{hori_joint_2017}. Given Whisper's large 50k vocabulary size and fixed sequence length, adding an extra projection layer poses memory challenges. To overcome this, we added two convolutional layers, each with a subsampling factor of two, along with an additional self-attention layer. 
Both the CTC head and the decoder are trained with timestamp tokens, and the CTC loss weight is set to 0.3.

For the ablation experiments, we used Whisper-medium.en, while the final model was trained with Whisper-large-v3-turbo. All models are trained with an overall batch size of 64~samples using bf16 precision and the AdamW optimizer~\cite{loshchilov2018decoupled}. The learning rate is set to $2\times 10^{-6}$, with a weight decay of $1\times 10^{-6}$, a linear decay scheduler, and 2k warm-up steps. The new parameters introduced by FDDT are trained with a learning rate of $2\times 10^{-4}$. By default, FDDT modules are inserted before all layers of the encoder with the diagonal constraint, meaning that only the diagonal values of the weight matrices are updated during training and can be non-zero. Unless otherwise stated, the CTC head undergoes an initial ``CTC preheating" phase, where it is trained on LibriSpeech for 10k steps, with the rest of the model being frozen. Afterwards, FDDT and CTC parameters are trained for a single epoch (FDDT preheating). Finally, the full model is trained for up to 50k steps, with early stopping set to a patience of 5 epochs. Most of the models typically converge within ten epochs. For the final evaluation, we always select the best-performing checkpoint based on the development set WER.

\subsection{Comparison to Baselines}
Table~\ref{tab:SOTA_comp} presents a comparison of the proposed method with various end-to-end and modular systems. The top section includes systems where no additional ground truth information about speaker identity or segmentation is provided, and the system must infer this information. For the proposed system, we utilized DiariZen~\cite{han2024leveraging}\footnote{\url{https://github.com/BUTSpeechFIT/DiariZen}} to condition our model.
The bottom section features models that directly or indirectly utilize ground truth segmentation information. It can be seen that our approach significantly outperforms the naive input masking baseline across all three datasets, largely due to its fine-tuning capabilities and robust handling of overlapped speech. Although our method does not achieve the top performance on the NOTSOFAR dataset, it surpasses baseline models, including the NOTSOFAR baseline~\cite{vinnikov24_interspeech} and the fine-tuned SURT model~\cite{raj23_surt2}. Furthermore, our approach delivers strong results on the AMI and Libri2Mix datasets, underscoring its effectiveness across diverse scenarios.

\subsection{Frame-Level Diarization Dependent Transformation}
\label{sec:ab_fddt}

\begin{table}[t]
    \centering
    \caption{%
        Analysis of different constraints applied to the FDDT parameters and methods used to initialize them evaluated with Whisper-medium.en on NOTSOFAR-1 eval-small. The FDDT parameters column specifies which parameters are used to condition the model. $\mathbf{W}_{diag} = diag(\mathbf{w})$, where $diag(\mathbf{w})$ is a diagonal matrix with elements of $\mathbf{w}$ in the diagonal.
    }\label{tab:FDDT_init}
    \small{\begin{tabular}{lccccc}
        \toprule
          & \multicolumn{3}{c}{Initilization Method}  \\
         FDDT parameters & Random & Identity & Suppressive \\
        \midrule

        \multirow{1}{*}{$\mathbf{b}$}
        & 28.4 & 28.0 & 28.0 \\

        \multirow{1}{*}{$\mathbf{W}_{diag}, \mathbf{b}$}
        & 129.4 & 27.3 & \textbf{26.7} \\

        \multirow{1}{*}{$\mathbf{W}, \mathbf{b}$}
        & 129.0 & 46.1 & 44.6 \\
        \bottomrule
    \end{tabular}}
\end{table}

To assess the impact of FDDT, we evaluated whether reducing the number of parameters in the additional modules affects the performance and examined the importance of proper initialization. As shown in Table~\ref{tab:FDDT_init}, using biases alone performs comparably to the combination of diagonal transformation matrices and biases, indicating that biasing frame-by-frame representations is sufficient to effectively focus the model on frames belonging to the same STNO class. The table also highlights that randomly initializing FDDT parameters is suboptimal and can significantly harm model performance, suggesting that suppressive initialization is preferable. Interestingly, using non-restricted weights, meaning full matrices without diagonal constraints leads to noticeable performance degradation.

\begin{table}[t]
    \centering
    \caption{%
        Effect of increasing the number of encoder layers, where FDDTs are applied on NOTSOFAR-1 eval-small with Whisper-medium.en (24 encoder layers).
    }\label{tab:FDDT_n_layers}
    \small{\begin{tabular}{lccccc}
        \toprule
        & &  \multicolumn{3}{c}{Initilization Method}  \\
        FDDT parameters & \# layers &   Random & Identity & Suppressive \\
        \midrule
       \multirow{3}{*}{$\mathbf{b}$}
        &   1  & 28.7 & 30.9 & 29.3 \\
        &   12 & 28.7 & 27.6 & 27.6  \\
        &   24 & 28.4 & 28.0 & 28.0 \\
       \midrule                            
        \multirow{3}{*}{$\mathbf{W}_{diag}, \mathbf{b}$} 
        &1  & 117.7 & 27.8 & 27.0 \\
        &12 & 118.9 & 27.4 & 27.1 \\
        &24 & 129.4 & 27.3 & \textbf{26.7} \\
     
        \bottomrule
    \end{tabular}}
\end{table}

Table~\ref{tab:FDDT_n_layers} demonstrates the effect of increasing the number of encoder layers where FDDTs are applied. The number of layers refers to how many layers, starting from the first encoder layer, are modified by FDDT parameters. For example, when only one layer is used, only the input to the Whisper encoder is modified.
The results indicate that even a single layer of bias-only parameters can achieve performance comparable to the best diagonal setup. However, using a random initialization with diagonal matrices, even for just the first layer, severely impacts the model's performance. Notably, the use of a diagonal setup with both $\mathbf{W}_{diag}$ and $\mathbf{b}$ across 24 layers yields the best performance, with the suppressive initialization method achieving the lowest WER of 26.7.

\begin{table}[b]
    \centering
    \caption{%
        The performance of FDDT given a reduction of information provided from diarization output on NOTSOFAR-1 eval-small with Whisper-medium.en. When employing the STNO mask, all frames are transformed, while with TNO, frames corresponding to the silence are left unchanged.
    }\label{tab:FDDT_info}
    \small{\begin{tabular}{ccccc}
        \toprule
         STNO & TNO & TN & T \\
        \midrule
        \textbf{26.7} & 30.0 & 28.7 & 34.8\\
        \bottomrule
    \end{tabular}}
\end{table}

Table~\ref{tab:FDDT_info} presents the performance of the FDDT method under different reductions of diarization information. 
It demonstrates that while the FDDT model still significantly outperforms input masking, some performance degradation is evident when comparing the STNO and T configurations. Interestingly, using three classes (TNO) results in worse performance than using (TN) classes only, even though the STNO and TNO configurations provide the same amount of information.

\begin{table}[t]
    \centering
    \caption{%
        Different sizes of training corpora affecting the performance of Whisper-medium.en. Tested on NOTSOFAR-1 Eval Small.
    }\label{tab:ab_data}
    \small{\begin{tabular}{ccc}
        \toprule
        NOTSOFAR-1 & + AMI & + Libri2Mix  \\
        \midrule
         26.7 & 25.6 & \textbf{24.8} \\        
        \bottomrule
    \end{tabular}}
\end{table}
\subsection{Scaling System With More Data And Parameters}
\label{sec:ab_scaling}
Table~\ref{tab:ab_data} explores the impact of incorporating additional training data on system performance. The results show that adding the AMI dataset improves the performance on the NOTSOFAR-1, highlighting the benefits of more real-world data. Furthermore, incorporating Libri2Mix data provides additional gains, raising the question of whether pretraining on synthetic data could offer even greater performance gains.

Finally, Table~\ref{tab:CTC_influence} provides a performance analysis across different model sizes, highlighting the improvements from employing an additional CTC head. Notably, incorporating a randomly initialized CTC head does not improve performance and leads to degradation. However, preheating the CTC head provides some improvements, especially for the small and medium models, even without joint decoding. Additionally, an FDDT preheating phase further enhances performance, with the best results observed in the large-v3 model.

 \begin{table}[ht]
    \centering
    \caption{%
        Influence of CTC head and size of the model evaluated on NOTSOFAR-1 eval-small.
    }\label{tab:CTC_influence}
    \small{
    \begin{tabular}{lcccc}
        \toprule
        &small.en & medium.en & large-v3 \\
        \midrule
        without CTC & 30.3 & 28.1 & 24.6  \\
        with CTC & 31.0 & 28.9 & 25.8 \\
        + CTC Preheating & 29.2 & 26.7 & 25.2 \\
        + FDDT Preheating & 30.3 & 27.4 & \textbf{24.5} \\
        \bottomrule
    \end{tabular}}
\end{table}

\section{Conclusions and Limitations}
\label{sec:conclusions}
In this study, we introduced FDDT into the Whisper model, making it a target-speaker ASR (TS-ASR) system conditioned on diarization output. We analyzed factors such as model size, FDDT placement, dataset size, and parameter initialization, evaluating the system on both real and synthetic datasets with consistently strong results. We also examined the impact of incorporating a CTC head, noting some improvements when it was preheated.

While the FDDT-based TS-ASR method performed effectively across diverse datasets, further validation on varied datasets, conditions, and languages is needed. We identified strengths and limitations, particularly regarding robustness against diarization errors, which need further attention for better real-world performance. Finally, the approach is extendable to other pre-trained ASR models, and a comparative analysis across ASR architectures could offer deeper insights into its adaptability and benefits.

\section*{Acknowledgement}
The work was supported by Ministry of Education, Youth and Sports of the Czech Republic (MoE) through the OP JAK project "Linguistics, Artificial Intelligence and Language and Speech Technologies: from Research to Applications" (ID:CZ.02.01.01/00/23\_020/0008518) and Brno Ph.D. Talent Scholarship Programme. Computing on IT4I supercomputer was supported by MoE through the e-INFRA CZ (ID:90254).

\bibliographystyle{IEEEtran}
\bibliography{references}

\end{document}